\documentclass[11pt]{article}
\pdfoutput=1
\usepackage{jheparxiv}

\usepackage[dvipsnames]{xcolor}

\usepackage{tikz-feynman}
\usepackage{amsmath}
\usepackage{amsfonts}
\usepackage{bm}
\usepackage{braket}

\usepackage[normalem]{ulem} %strikethrough text

\usepackage{hyperref}
\hypersetup{colorlinks= true, linkcolor=blue, citecolor=red}

\newcommand{\be}{\begin{equation}}
\newcommand{\ee}{\end{equation}}

\newcommand{\kk}{\ell}

\newcommand{\pd}{\partial}
\newcommand{\ud}{\mathrm{d}}
\newcommand{\dhat}{\hat{\ud}}
\newcommand{\deltahat}{\hat{\delta}}

\newcommand{\DCto}{\underset{\text{D.C.}}{\to}}

\newcommand{\cc}{\text{c.c.}}

\newcommand{\sfT}{{\mathsf T}}

\numberwithin{equation}{section}

\newcommand{\LCperp}{{\scriptscriptstyle \perp}}

\usepackage{mathrsfs}

\newcommand{\Abg}{\mathscr{A}}
\newcommand{\Abgpert}{\mathscr{A}}
\newcommand{\Avac}{\mathcal{A}}
\newcommand{\Astr}{\mathcal{A}}

\newcommand{\Mbg}{\mathscr{M}}
\newcommand{\Mbgpert}{\mathscr{M}}
\newcommand{\Mvac}{\mathcal{M}}
\newcommand{\Mstr}{\mathcal{M}}

\newcommand{\C}{C} %%decomposition constants

\title{Coherent states, background fields, and double copy}
\author{}
\date{}

\begin{document}

\author{Anton Ilderton}
\emailAdd{anton.ilderton@ed.ac.uk}

\author{and William Lindved}
\emailAdd{william.lindved@ed.ac.uk}

\affiliation{Higgs Centre, School of Physics \& Astronomy, University of Edinburgh, EH9 3FD, Scotland, UK}

\abstract{
We show that scattering amplitudes on any gauge theory background admitting a coherent state description double copy to amplitudes in a curved spacetime. The metric of the spacetime is built from the gauge background using a notion of classical double copy which emerges naturally at the amplitude level. In the self-dual sector this map relates backgrounds which are exact vacuum solutions in gauge theory and gravity.
}
\maketitle

\section{Introduction}

The double copy has become a valuable tool for computing amplitudes in gravitational theories, constructing them from simpler amplitudes in gauge theory~\cite{Kawai:1985xq,Bern:2008qj,Bern:2010ue,Bern:2010yg,Cachazo:2013gna,Cachazo:2013hca}. 
For reviews see~\cite{Bern:2019prr,Bern:2022wqg,Adamo:2022dcm}.
While the double copy of amplitudes on flat space is well understood, the question remains of whether a double copy relation exists in the presence of background gauge fields and in curved spacetimes.
Work on ``background'' double copy falls broadly into one of two categories~\cite{Bahjat-Abbas:2017htu}. Gauge theory amplitudes in curved spacetime could
double copy to gravitational amplitudes in the same spacetime, as explored in e.g.~\cite{Albayrak:2020fyp, Zhou:2021gnu, Diwakar:2021juk, Cheung:2022pdk, Herderschee:2022ntr, Drummond:2022dxd, Lee:2022fgr, Lipstein:2023pih, Mei:2023jkb, Brown:2023zxm, Liang:2023zxo, CarrilloGonzalez:2024sto,Beetar:2024ptv, Alday:2025bjp}.
Another possibility is that gauge theory amplitudes on flat space, but in the presence of a gauge background, double copy to gravitational amplitudes in a curved spacetime with metric somehow connected to the initial gauge background. For examples of amplitudes and observables which follow this construction see e.g.~\cite{Luna:2016hge,Adamo:2017nia,Emond:2020lwi,Gonzo:2021drq,Ilderton:2024oly}.

Turning to classical results, many examples have been found of metrics that can be generated by `squaring' some part of
a gauge field. This is known as \textit{classical} double copy. There are a variety of approaches~\cite{Monteiro:2014cda,Luna:2015paa,Goldberger:2017vcg,Luna:2018dpt,Kim:2019jwm,Alkac:2021bav,Adamo:2021dfg,Chacon:2021wbr,Chacon:2021lox,Shi:2021qsb,Comberiati:2022cpm,Kent:2024mow}, for a review see~\cite{White:2024pve}, but questions remain about the precise connection of any classical double copy to its amplitudes-based namesake~\cite{Luna:2016due,Goldberger:2016iau,Bautista:2019tdr}. For example, how does a local double copy in position space arise from the momentum-space formulation at the amplitude level~\cite{Luna:2022dxo}? If a metric and gauge background are related by a classical double copy, are amplitudes on those backgrounds also related by some double copy? Below, we will identify a definition of classical double copy for which the answer to this latter question is yes, for backgrounds admitting a coherent state description.

Coherent states have received increased attention in the amplitudes community in recent years, not least because the classical limit of the two-body problem produces a final radiation state which is effectively a coherent state of gravitons~\cite{Ciafaloni:2018uwe,Cristofoli:2021vyo, Britto:2021pud, Cristofoli:2021jas,DiVecchia:2022nna,DiVecchia:2022piu}. Other applications include, for example, investigating the impact of large gauge effects on amplitudes~\cite{Cristofoli:2022phh,Elkhidir:2024izo}, and modelling absorptive processes~\cite{Endlich:2016jgc,Ilderton:2017xbj,Aoude:2023fdm,Aoude:2024jxd,Adamo:2025vzv}. As coherent states give non-zero expectation values to gauge fields and metric perturbations, they are intimately related to classical backgrounds. {In fact}, amplitudes containing coherent states can equivalently be computed in background field perturbation theory, where the field operators are replaced by dynamical quantum fluctuations around the classical background defined by the coherent states. 

Our aim here is to learn about double copy on those background fields which can be represented by coherent states. {Since} coherent state amplitudes can be expanded in terms of ordinary number state amplitudes, it suggests that we can make use of known double copy relations for \emph{vacuum} amplitudes to learn about double copy on backgrounds. Indeed, our guiding principle will be to make maximal use of known relations whilst introducing minimal additional structure. Of course, there exists variety of formulations of double copy for amplitudes on flat space. These are not necessarily identical maps, and can have different ranges of applicability. In what follows we will not favour any particular double copy scheme, instead focussing on extending the double copy, in whatever form, to amplitudes on coherent state backgrounds.

Our starting point will be scattering in flat space {between asymptotic states which} contain coherent states of on-shell gluons/gravitons. Expanding these amplitudes in the number of coherent state particles, we will see that the gauge and gravity amplitudes have very similar structure at each order in the expansion. This will lead us to propose a perturbative double copy scheme in which \textit{vacuum} amplitudes are double copied in the standard way while all other quantities are left undisturbed. Under this map coherent state amplitudes in gauge theory turn into coherent state amplitudes in some axio-dilaton gravity theory. In the background field picture, this leads naturally to a notion of classical double copy relating the gauge and gravity backgrounds -- this {was not a requirement of our map}, but rather a direct consequence.

This paper is organised as follows. In Sec.~\ref{sec:coherent} we discuss scattering with coherent states of gluons/gravitons {and its equivalent description in terms of amplitudes on background fields}, with an emphasis on the diagrams contributing in both pictures. This is in preparation for the discussion of the double copy of coherent state amplitudes, which is the subject of Sec.~\ref{sec:DC}. Here, working first in perturbation theory, we use vacuum double copy to construct a double copy {on backgrounds}. We then present a non-perturbative map which reproduces the perturbative results. In Sec.~\ref{sec:classicalDC} we turn to classical double copy. We give several explicit examples of the classical double copy map arising from our amplitudes results, including plane waves, {Kerr-Schild for self-dual fields}, and double Kerr-Schild. 
We also compare and contrast with other notions of classical double copy.
We conclude in Sec.~\ref{sec:concs}.
%
%%%%%%%%%%
\subsection*{Notation and conventions}
An integral with subscript is an integral over on-shell massless modes, i.e.~
\be
    \int_k \equiv \int \frac{\ud^3\mathbf{k}}{(2\pi)^3 2|\mathbf{k}|} \;.
\ee
The on-shell delta function $\delta_{\text{o.s.}}(k-k')$ obeys
\be
    \int_k \delta_\text{o.s.}(k-k') = 1 \;.
\ee
Factors of $2\pi$ absorbed into delta functions and differentials are indicated by
\begin{equation}
 \dhat^nk \equiv \frac{\ud^n k}{(2\pi)^n} \;, \quad \deltahat^n(k) \equiv (2\pi)^n \delta^n(k) \;.  
\end{equation}
%%%%%%%%%%%%%%
\section{Coherent states and amplitudes}\label{sec:coherent}
%%%%%%%%%%%%%%
\subsection{Perturbation theory with coherent states}
Beginning in gauge theory, we consider matrix elements of the form
\be\label{starting-point}
    \Abg\equiv \bra{\alpha',\text{out}} T \ket{\alpha,\text{in}} \;,
\ee
in which $\alpha$ and $\alpha'$ represent two coherent states of gluons, while `in' and `out' represent number states of massive particles and possibly other gluons.  To make this explicit, we expand the gauge field as usual as
\be\label{gauge-mode-expansion}
    \hat{A}^a_{\mu}(x) = \int_k a^a_s(k) \varepsilon_{\mu}^s(k)e^{-ik\cdot x} + a^{a\dagger}_s(k) {\bar\varepsilon}_{\mu}^s(k)e^{ik\cdot x} \;,
\ee
where $a$ labels the colour, $\varepsilon_{\mu}^s(k)$ are a basis of polarisation vectors %for helicity $s$
(repeated colour and spin indices are implicitly summed over), and the mode operators obey
\be
    [a^a_s(k),a^{b\dagger}_{s'}(k')] = \delta_{ab}\delta_{ss'}\delta_{\text{o.s.}}(k-k')  \;.
\ee
Number states of gluons are, as usual, given by products of $a^{a\dagger}_s$ acting on the vacuum. Our approach will be based on scattering of coherent states built from asymptotically free gluons. We construct these by acting on the vacuum with the displacement operator\footnote{There is no operator ordering ambiguity in defining the states in (\ref{starting-point}), assuming the number-states have negligible phase-space overlap with the coherent states, as is the case for scattering of well-separated constituents.}, $\ket{\alpha} = D(\alpha) \ket{0}$, where
\be\label{def-displacement}
\begin{split}
    D(\alpha)
    &= e^{\int_k a^{a\dagger}_s(k)\alpha^a_s(k) - a^a_s(k){\bar \alpha}^a_s(k)} 
    = e^{\int_k a^{a\dagger}_s(k)\alpha^a_s(k)}e^{- \int_k a^a_s(k){\bar \alpha}^a_s(k)}e^{-\frac{1}{2}\int_k {\bar \alpha}^a_s(k)\alpha^a_s(k)}\;,
\end{split}
\ee
and we use $\alpha$ as shorthand for the set of complex profile functions $\alpha^a_s(k)$.
The key property of such coherent states is that, unlike number states, they generate non-zero expectation values for the asymptotic gauge field (\ref{gauge-mode-expansion}):
\be\label{gauge-pert-expect-val}
    \bra{\alpha} {\hat A}^a_{\mu}(x) \ket{\alpha} = \int_k \alpha^a_s(k) \varepsilon_{\mu}^s(k)e^{-ik\cdot x} + {\bar \alpha}^a_s(k) {\bar\varepsilon}_{\mu}^s(k)e^{ik\cdot x} \;.
\ee
A suitable starting point for a perturbative expansion of the amplitude (\ref{starting-point}) is given by simply writing out all coherent states explicitly using (\ref{def-displacement}), thus
\begin{equation}\label{gauge-amp-unexpanded}
    \Abg  = e^{- \tfrac12
    \int_k|\alpha_s^a|^2 + |\alpha_s^{\prime a}|^2 } \bra{\text{out}} e^{
    \int_k a^a_s(k)\bar\alpha^{\prime a}_s(k)}   T e^{
    \int_k a^{a\dagger}_s(k)\alpha^a_s(k)}\ket{\text{in}} \;.
\end{equation}
We will consider the expansion of~(\ref{gauge-amp-unexpanded}) in \emph{the total number of particles contributing from the coherent states}, which we call $n$.
We thus expand the exponentials between the bra and ket, but not the prefactor (we will revisit this below), writing
\begin{equation}\label{Abg-from-Astr}
     \Abg=  e^{- \tfrac12
    \int_k|\alpha_s^a|^2 + |\alpha_s^{\prime a}|^2 } \sum_n \Abgpert_{(n)} \;.
\end{equation}
To illustrate this and introduce notation, the lowest order terms in the sum above are
\begin{equation}\label{calM-expanded-leading}
    \Abgpert_{(0)} = \bra{\text{out}} T \ket{\text{in}}  \;,
    \qquad
    \Abgpert_{(1)}=
    \int_k \bar \alpha^{\prime a}_s(k) \bra{\text{out}, k^{a,s}} T\ket{\text{in}} + \alpha_s^a(k) \bra{\text{out}} T \ket{k^{a,s}, \text{in}} \;.
\end{equation}
The $n=0$ term is independent of the coherent states. Compared to this the $n=1$ terms contain one additional gluon in either the initial or final state, with momentum $k$ and quantum numbers $a$, $s$. The coherent state profiles act essentially as (non-normalised) wavepackets for these particles. This underlines in particular that at fixed $n$ one can have terms with different powers of $\alpha$ and $\alpha'$.

Clearly the matrix elements in (\ref{calM-expanded-leading}), and those encountered at higher $n$, are ordinary amplitudes for number states in vacuum (or simply `vacuum amplitudes' from here on). In an attempt to minimise clutter, we adopt the following compact notation: for vacuum amplitudes in which the initial/final coherent state contributes a gluon of momentum $k_j$/$k_i$ and quantum numbers $\{a_j, s_j\}$/\{$a_i,s_i\}$, we write
%
%Clearly the matrix elements in (\ref{calM-expanded-leading}), and those encountered at higher $n$, are ordinary matrix elements of number states in vacuum. In an attempt to minimise clutter, we adopt the following compact notation: for matrix elements in which the initial/final coherent state contributes a gluon of momentum $k_j$/$k_i$ and quantum numbers $\{a_j, s_j\}$/\{$a_i,s_i\}$, we write
%
\be\label{gauge-amp-def-vacuum}
        \Avac(\ldots {\bar i} \ldots j \ldots) \equiv \bra{\text{out},\ldots k_{i}^{a_i,s_i}\ldots} T \ket{\ldots k_{j}^{a_j,s_j}\ldots , \text{in}} \;.
\ee
It is worth noting that~(\ref{gauge-amp-def-vacuum}) is, in principle, a full, non-perturbative vacuum amplitude.
%It is worth noting that~(\ref{gauge-amp-def-vacuum}) is, in principle, a full non-perturbative flat space amplitude.
With this, we can write the higher-$n$ terms $\Abgpert_{(n)}$ appearing in the expansion (\ref{Abg-from-Astr}) as
\begin{equation}\label{gauge-amp-expanded}
\begin{split}
   \Abgpert_{(n)} = \sum_r\frac{1}{r! (n - r)!}
   \int_{k_1 \dots k_n}  { \sum_{\{ a_i, s_j \} } }
   \bar \alpha_{s_1}^{\prime a_1}(k_1) \dots &\bar \alpha_{s_r}^{\prime a_r}(k_r)  \alpha^{a_{r+1}}_{s_{r+1}}(k_{r+1}) \dots \alpha_{s_n}^{a_n}(k_n)\times \\
   &\times \Avac(\bar 1, \ldots, \bar r, r+1, \ldots n) \;,
\end{split}
\end{equation}
in which $r$ runs over the number of \emph{outgoing} gluons. We have written out explicitly the sum over colour and spin, as the indices of the external states  have been absorbed into the definition of $\Avac$.

\begin{figure}[t!]
\begin{equation*}
        \begin{tikzpicture}[baseline=(o.base)]
			\begin{feynman}[small]
				\vertex (o) at (0,0);
                \vertex[left = 0.3cm of o] (vbg);
                \vertex[right = 0.3cm of o] (v2);
                \vertex[left = 0.3 cm of vbg] (v1);
                \vertex[right = 0.5cm of v2] (fermut);
                \vertex[left = 0.5 cm of v1] (fermin);
                \vertex[above = 0.8 cm of fermin] (gluin);
                \vertex[above = 0.8cm of fermut] (gluut);
                \vertex[below = 0.8 cm of fermin] (bgin);
				\diagram* {
					(fermin) -- [fermion] (fermut),
                    (gluin) -- [gluon] (v1),
                    (gluut) -- [gluon] (v2),
                    (bgin) -- [gluon] (vbg),
				};
                \node[anchor = north east] at ($(bgin) - (-0.14, -0.14)$) {$\times$};
			\end{feynman}
		\end{tikzpicture}  \qquad
        \begin{tikzpicture}[baseline=(oglu.base)]
			\begin{feynman}[small]
				\vertex (o) at (0,0);
                \vertex[left = 1cm of o] (fermin);
                \vertex[right = 1 cm of o] (fermut);
                \vertex[above = 1 cm of o] (oglu);
                \vertex[left = 1 cm of oglu] (gluin);
                \vertex[right = 1cm of oglu] (gluut);
                \vertex[above = 1.2 cm of oglu] (bgc);
                \vertex[above left = 0.6cm of bgc] (bgi1);
                \vertex[below left = 0.6cm of bgc] (bgi2);
                \vertex[right  = 0.7cm of bgc] (bgu);
				\diagram* {
					(fermin) -- [gluon] (o) -- [gluon] (fermut),
                    (gluin) -- [fermion] (oglu) -- [fermion] (gluut),
                    (o) -- [gluon] (oglu),
                    (bgi1) -- [gluon] (bgc) -- [gluon] (bgu),
                    (bgi2) -- [gluon] (bgc),
				};
                \node[anchor = west] at ($(bgu) - (0.1, 0)$) {$\times$};
                \node[anchor = north east] at ($(bgi2) - (-0.14, -0.14)$) {$\times$};
                \node[anchor = south east] at ($(bgi1) - (-0.14, 0.14)$) {$\times$};
			\end{feynman}
                                        \node[draw] at (2, 3) {SMC};
            	\begin{feynman}[small]
				\vertex (o) at (3,1);
                \vertex[left = 0.4 cm of o] (v1);
                \vertex[left = 0.4 of v1] (fermin);
                \vertex[right = 0.4 cm of o] (v2);
                \vertex[right = 0.4cm of v2] (fermut);
                \vertex[below = 0.7cm of fermin] (gluin);
                \vertex[below = 0.7cm of fermut] (bgut);
                \vertex[above = 1 cm of o] (bgc);
                \vertex[left = 0.7cm of bgc] (bgi);
                \vertex[right  = 0.7cm of bgc] (gluut);
				\diagram* {
					(fermin) -- [fermion] (fermut),
                    (gluin) -- [gluon] (v1),
                    (bgut) -- [gluon] (v2),
                    (bgi) -- [gluon] (gluut),
				};
                \node[anchor = north west] at ($(bgut) - (0.14, -0.14)$) {$\times$};
                \node[anchor = north east] at ($(gluin) - (-0.14, -0.14)$) {$\times$};
			\end{feynman}           
		\end{tikzpicture}  \qquad
        \begin{tikzpicture}[baseline=(o.base)]
			\begin{feynman}[small]
				\vertex (o) at (0,0);
                \vertex[left = 0.4 cm of o] (v1);
                \vertex[left = 0.4 of v1] (fermin);
                \vertex[right = 0.4 cm of o] (v2);
                \vertex[right = 0.4cm of v2] (fermut);
                \vertex[below = 0.7cm of fermin] (gluin);
                \vertex[below = 0.7cm of fermut] (bgut);
                \vertex[above = 1 cm of o] (bgc);
                \vertex[left = 0.7cm of bgc] (bgi);
                \vertex[right  = 0.7cm of bgc] (gluut);
				\diagram* {
					(fermin) -- [fermion] (fermut),
                    (gluin) -- [gluon] (v1),
                    (bgut) -- [gluon] (v2),
                    (bgi) -- [gluon] (gluut),
				};
            \node[anchor = west] at ($(gluut) - (0.2, 0)$) {$\times$};
            \node[anchor = north west] at ($(bgut) - (0.14, -0.14)$) {$\times$};
			\end{feynman}
            \draw[thick] (1.6,-2) -- (1.6,2);
		\end{tikzpicture} \quad
        \begin{tikzpicture}[baseline=(o.base)]
			\begin{feynman}[small]
				\vertex (o) at (0,0);
                \vertex[left = 0.4 cm of o] (v1);
                \vertex[left = 0.4 of v1] (fermin);
                \vertex[right = 0.4 cm of o] (v2);
                \vertex[right = 0.4cm of v2] (fermut);
                \vertex[below = 0.7cm of fermin] (gluin);
                \vertex[below = 0.7cm of fermut] (bgut);
                \vertex[above = 1 cm of o] (bgc);
                \vertex[left = 0.7cm of bgc] (bgi);
                \vertex[right  = 0.7cm of bgc] (gluut);
				\diagram* {
					(fermin) -- [fermion] (fermut),
                    (gluin) -- [gluon] (v1),
                    (bgut) -- [gluon] (v2),
                    (bgi) -- [gluon ] (gluut),
				};
                \node[anchor = east] at ($(bgi) + (0.1, 0)$) {$\times$};
                \node[anchor = west] at ($(gluut) - (0.2, 0)$) {$\times$};
			\end{feynman}
		\end{tikzpicture}
\end{equation*}
\caption{\label{fig:explain}  A selection of diagrams in $2\to 2$ Compton scattering of quarks and gluons. External gluons belonging to the coherent states are marked with a cross. The rightmost diagram is not SMC and does not need to be computed, as per the final line of~(\ref{SMC-first-def}).}
\end{figure}
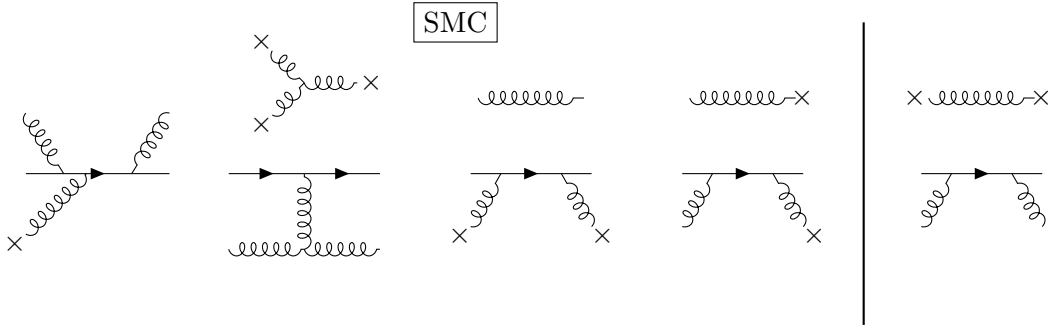

As written, the vacuum amplitudes $\Avac$ contain disconnected diagrams, many of which we might normally ignore, {that} we need to keep track of when we relate to the background field picture of coherent states, below. These are diagrams in which gluons from the initial coherent state run straight through, without interacting, and contract with gluons in the final coherent state, giving a factor
%
%As written, the vacuum matrix elements $\Avac$ contain disconnected diagrams, many of which we might normally ignore, {that} we need to keep track of when we relate to the background field picture of coherent states, below. These are diagrams in which gluons from the initial coherent state run straight through, without interacting, and contract with gluons in the final coherent state, giving a factor
%
\iffalse
\begin{equation}\label{propagator-grav}
    		\begin{tikzpicture}[baseline=(o.base)]
			\begin{feynman}
				\vertex (o) at (0,0){$P$};
				\vertex[left= 2.5cm of o] (s){$P'$};			
				\diagram* {
					(s) -- [photon] (o),
				};
			\end{feynman}
		\end{tikzpicture} 
  = \delta_P^{P'} \delta_\text{o.s.}( k' - k),
\end{equation}
\fi
\begin{equation}\label{propagator-gauge}
    		\begin{tikzpicture}[baseline=(o.base)]
			\begin{feynman}
				\vertex (o) at (0,0){$k_2^{ a,s}$};
				\vertex[left= 2.5cm of o] (s){$k_1^{b,\sigma}$};			
				\diagram* {
					(s) -- [gluon] (o),
				};
			\end{feynman}
		\end{tikzpicture} 
  = \delta_{ab}\delta^{s\sigma}\delta_\text{o.s.}( k_2 - k_1).
\end{equation}
The overall contribution of such diagrams is {particularly simple}. Consider for example that at second order, $\Abgpert_{(2)}$ contains a contribution from a disconnected line (\ref{propagator-gauge}) multiplying the coherent-state-independent amplitude $\Abgpert_{(0)}$; writing a `blob' for the latter, the relevant contribution is
%
\iffalse
\begin{equation}\label{mult-factor}
    \sum_{P_1P_2}\int_{k_1k_2} \bar \beta'_{P_1}(k_1) \beta_{P_2}(k_2)             \begin{tikzpicture}[baseline=(o.base)]
			\begin{feynman}[]
				\vertex (o) at (0,0);
                \vertex[above = 0.4 cm of o] (o1);
				\vertex[blob, below= 0.1cm of o] (s){};
                \vertex[left = 1cm of o1] (g1){$P_1$};
                \vertex[right = 1cm of o1] (g2){$P_2$};
				\diagram* {
					(g1) -- [photon] (g2),
				};
			\end{feynman}
		\end{tikzpicture}
        =  \;
        \begin{tikzpicture}[baseline=(o.base)]
			\begin{feynman}
				\vertex[blob] (o) at (0,0){};
			\end{feynman}
		\end{tikzpicture}
        \;\;
        \sum_{P_1} \int_{k_1} \bar \beta'_{P_1} (k_1) \beta_{P_1} (k_1)     	.	    
\end{equation}
\fi
%
\begin{equation}\label{mult-factor}
    \int_{k_1k_2} {\bar\alpha}^{\prime a_1}_{s_1}(k_1) \alpha_{s_2}^{a_2}(k_2)
    \begin{tikzpicture}[baseline=(o.base)]
			\begin{feynman}[]
				\vertex (o) at (0,0);
                \vertex[above = 0.4 cm of o] (o1);
				\vertex[blob, below= 0.1cm of o] (s){};
                \vertex[left = 1cm of o1] (g1){$k^{a_1,s_1}_1$};
                \vertex[right = 1cm of o1] (g2){$k^{a_2,s_2}_2$};
				\diagram* {
					(g1) -- [gluon] (g2),
				};
			\end{feynman}
		\end{tikzpicture}
        =  \;
        \begin{tikzpicture}[baseline=(o.base)]
			\begin{feynman}
				\vertex[blob] (o) at (0,0){};
			\end{feynman}
		\end{tikzpicture}
        \;\;
    \int_{k} {\bar \alpha}^{\prime a}_{s} (k) \alpha_{s}^{a} (k)     	.	    
\end{equation}
Clearly, the higher order terms $\Abgpert_{(n)}$ all contain diagrams in which $\Abgpert_{(0)}$ is `dressed' with some number of propagators; combining all of them, the {profile-dependent factor} in (\ref{mult-factor}) exponentiates. Moreover, this feature applies not only to $\Abgpert_{(0)}$, but the effect of computing all diagrams in~(\ref{gauge-amp-unexpanded}) containing at least one propagator~(\ref{propagator-gauge}) connecting gluons in the coherent states is simply to provide an overall  factor
\be
	\exp \int_{k} {\bar \alpha}^{\prime a}_{s} (k) \alpha^a_{s}(k) \;.
\ee
This means that to compute the full scattering amplitude between coherent states we \emph{do} need to consider disconnected contributions, but in practice we need not evaluate \emph{all} of them -- it is enough to consider `slightly more connected' (SMC) diagrams, meaning those in which all free propagators (\ref{propagator-gauge}) \emph{between coherent state particles} are stripped away. {So we may write}
\begin{equation}\label{SMC-first-def}
\begin{split}
    \Abg  &= e^{- \tfrac12
    \int_k|\alpha_s^a|^2 + |\alpha_s^{\prime a}|^2 } \bra{\text{out}} e^{
    \int_k a^a_s(k)\bar\alpha^{\prime a}_s(k)}   T e^{
    \int_k a^{a\dagger}_s(k)\alpha^a_s(k)}\ket{\text{in}} \bigg|_\text{all diagrams} \\
    &= e^W 
    \bra{\text{out}} e^{
    \int_k a^a_s(k)\bar\alpha^{\prime a}_s(k)}   T e^{
    \int_k a^{a\dagger}_s(k)\alpha^a_s(k)}\ket{\text{in}}\bigg|_\text{SMC diagrams}\\
    & = e^{W} \sum_n \Abgpert_{(n)} \bigg|_\text{SMC diagrams} \;,
\end{split}
\end{equation}
where the exponent $W$ is
\be\label{W-definition}
    W =  - \frac12
    \int_k|\alpha_s^a|^2 - \frac12
    \int_k |\alpha_s^{\prime a}|^2 
    +\int_{k} {\bar \alpha}^{\prime a}_{s} (k) \alpha^a_{s}(k)\;,
\ee
which has both real and imaginary parts. The SMC diagrams still contain disconnected pieces, {including some which only connect gluons} in the initial and final coherent states, but through interactions --  this is illustrated in Fig.~\ref{fig:explain}.
Our next objective 
is to understand these structures in the context of scattering on a background gauge field, which is an alternative way of thinking about coherent states.

%%%%%%%%%%%%%%%%%%%%%%%%%%%%
\subsection{Perturbation theory in background fields}
%%%%%%%%%%%%%%%%%%%%%%%%%%%%
%
A key property of the displacement operator which creates coherent states is that it displaces mode operators by a $\mathbb{C}$-number $\alpha$, i.e.
\be
    D^\dagger(\alpha) a^a_s(k) D(\alpha) = a_s^a(k) + \alpha_s^a(k) \;.
\ee
With this we return to our initial matrix element written in the form (\ref{gauge-amp-unexpanded}). We can think of the $S$-matrix, or $T$-matrix, as a function of the gluon ladder operators, writing (suppressing quantum numbers for a moment) $T\equiv T[a,a^\dagger]$.  Observe then that commuting the exponential on the right of  (\ref{gauge-amp-unexpanded}) past the $T$-matrix shifts $a(k)\to a(k) + \alpha(k)$ but leaves $a^\dagger(k)$ untouched, i.e.~
\be\label{translate-1}
T[a,a^\dagger] e^{%\sum_P
\int_k a^\dagger(k)\alpha(k)} = e^{%\sum_P
\int_k a^\dagger(k)\alpha(k)} T[a+\alpha,a^\dagger] \;.
\ee
To generate a shift also in $a^\dagger$ we take the leftmost exponential operator in (\ref{gauge-amp-unexpanded}), and move it to the right. Since this means commuting past the exponential in (\ref{translate-1}) as well, we generate an additional constant factor via the standard BCH formula, arriving at
\be\label{messy}
    \Abg = e^{W}\bra{\text{out}} e^{\int_k a^{a\dagger}_s(k)\alpha_s^a(k)}
    T[a+\alpha,a^\dagger + {\bar \alpha'}]
    e^{\int_k a^a_s(k)\bar\alpha^{\prime a}_s(k)}
    \ket{\text{in}}\;,
\ee
in which $W$ is precisely as in (\ref{W-definition}). The $T$-matrix in (\ref{messy}), with its shifted mode operators, is nothing but the $T$-matrix in the presence of a background gauge field constructed from the coherent state profiles as~\cite{Frantz:1965,Kibble:1965zza}
\be\label{background-gauge-definition}
\begin{split}
    A^{a}_{\mu}(x) = \int_k \alpha^a_s(k) \varepsilon_{\mu}^{s}(k)e^{-ik\cdot x} + {\bar \alpha}^{\prime a}_s(k) {\bar\varepsilon}_{\mu}^s(k)e^{ik\cdot x} \equiv \int\!\hat{\ud}^4 k\, A_\mu^a(k) e^{-ik\cdot x} \;,
\end{split}
\ee
where $A_\mu^a(k)$ is defined to contain the usual on-shell delta function $\delta(k^2)\theta(k_0)$. In terms of the underlying interaction Lagrangian, one translates the gauge field by the classical field (\ref{background-gauge-definition}). This background solves the \emph{linearised} Yang-Mills equations. Its incoming/positive energy modes are determined by $\alpha$, while the outgoing/negative energy modes are determined by $\alpha'$.
{If $\alpha'=\alpha$ the corresponding background} is real, but in general $A^a_\mu(x)$ will be complex-valued~\cite{Zwanziger:1973if}.
Physically we may then interpret certain complex backgrounds as an effective description of scattering between different coherent states.

The remaining exponential operators in (\ref{messy}) modify the asymptotic states. The rightmost exponential, for example, will remove gluons from the number state $\ket{\text{in}}$ if they have kinematic overlap with $\alpha'$, turning the initial state (hence the amplitude) into a sum over states (amplitudes) of lower gluon number, multiplied by factors of $\alpha'$. 
{These terms correspond to disconnected diagrams in the coherent state picture, namely those containing lines which connect in-state gluons directly to $\alpha'$  as illustrated in the fourth diagram of Fig.~\ref{fig:explain}. Assuming that the in-state gluons are kinematically distinct from the final coherent state these terms vanish\footnote{For this reason they are typically not included in background field calculations, though for a recent discussion see~\cite{Ilderton:2017xbj}.}, and we may drop the exponentials as we do below. (The same applies for out-state gluons connecting to $\alpha$.)}
An example case in which these terms are automatically absent is two-massive-particle scattering with coherent emission of radiation.
In the general case the exponentials should be kept, and it is straightforward to do so.

With this we arrive at the long-known equivalence between scattering in coherent states and in background fields,
\be\label{two-pictures}
        \Abg \equiv \bra{\alpha',\text{out}} T \ket{\alpha,\text{in}}
        \leftrightarrow
        e^{W}\bra{\text{out}}T^\text{{b.g.}}
    \ket{\text{in}} \;,
\ee
where $T^\text{b.g.}$ is shorthand for the shifted transition matrix from~(\ref{messy}).
We can examine the connection between coherent states and background fields more explicitly, beginning with the $n=1$ matrix element (\ref{calM-expanded-leading}) in the coherent state picture,
\be\label{spin-strip-intro}
     \Abg_{(1)} = %\sum_{a,s} 
     \int_{k}  \bar \alpha^{\prime a}_{s} \Avac(\bar1) + \alpha^a_{s} \Avac(1) \;.
\ee
We make the momentum-conserving delta functions in the vacuum amplitudes explicit, and strip the labelled gluon polarisations from the exposed vacuum amplitudes, writing
%
%We make the momentum-conserving delta functions in the vacuum matrix elements explicit, and strip the labelled gluon polarisations from the exposed vacuum amplitudes, writing
%
\begin{equation}
\begin{split}
\Avac(1) &\equiv  \varepsilon^\mu_s(k) \, \deltahat^4(\Delta Q - k) \, {\Astr_\mu^a(-k)} \;, \\
        \Avac({\bar 1}) &=  {\bar\varepsilon}^\mu_s(k) \, \deltahat^4(\Delta Q + k) \, {\Astr_\mu^a(+k)} \;,
\end{split}
\end{equation}
in which $\Delta Q \equiv Q_\text{out} - Q_\text{in}$ is the difference between the total momenta of the states $\ket{\text{out}}$ and $\ket{\text{in}}$. We refer to $\Astr_\mu^a$ as a \textit{spin-stripped amplitude}.
Switching perspectives to the background field picture, we associate the stripped polarisation vectors not with the vacuum amplitude but with the profile functions -- this essentially replaces external lines in the vacuum amplitudes with insertions of the background field (\ref{background-gauge-definition}). It is easily checked that
\begin{equation}\label{gauge-amp-coherent-bgft-pert}
\begin{split}
    \Abgpert_{(1)} &= \int \dhat^4 k \, \deltahat^4(\Delta Q {-k}) \, A_a^\mu(k) \,   \Astr_\mu^a(-k) \;.
\end{split}
\end{equation}
A similar treatment allows us to generalise~(\ref{gauge-amp-coherent-bgft-pert}) to all $n$. The spin-stripped $n$-point amplitude is defined in complete analogy to (\ref{spin-strip-intro}), e.g.
\begin{equation}
    \Avac({\bar 1}, \ldots, n) \equiv  {\bar\varepsilon}_{s_1}^{\mu_1} \dots \varepsilon_{s_n}^{\mu_n} \, \deltahat^4 \bigg(\Delta Q {+ k_1 +\ldots - k_n}
    %\sum_i k_i
    \bigg) \, \Astr^{a_1 \dots a_n}_{\mu_1 \dots \mu_n} (+k_1, \ldots, -k_n) \;.
\end{equation}
The permutation invariance of amplitudes, together with the fact that all insertions of the background $A_\mu^a(k)$ are identical,  allows the sum over $r$ in (\ref{gauge-amp-expanded}) to be performed. This yields
\begin{equation}\label{gauge-amp-coherent-bgft}
    \Abgpert_{(n)} = \frac{1}{n!} %\sum_{a_1 \dots a_n}
    \int \dhat^4 k_1 \dots \dhat^4 k_n \,
    \deltahat^4\bigg(\Delta Q {- \sum_i k_i} \bigg)
    A_{a_1}^{\mu_1} (k_1) \dots A_{a_n}^{\mu_n} (k_n) \Astr_{\mu_1 \dots \mu_n}^{a_1 \dots a_n} (-k_1, \ldots, -k_n) \;.
\end{equation}
which has the expected form of a background-field amplitude.

\subsection*{Gravity}
The story in gravity is largely the same. 
We expand the metric perturbation $\mathrm{h}_{\mu\nu}(x)$ as usual as  
\be\label{graviton-mode-expansion}
    \mathrm{h}_{\mu\nu}(x) = \int_k a_s(k) \varepsilon_{\mu\nu}^s(k)e^{-ik\cdot x} + a^\dagger_s(k) {\bar\varepsilon}_{\mu\nu}^s(k)e^{ik\cdot x} \;,
\ee
where $s$ labels some basis of graviton spin states with corresponding polarisation tensors $\varepsilon_{\mu\nu}^s(k)$, and the mode operators obey
\be
    [a_s(k),a^\dagger_{s'}(k')] = \delta_{\text{o.s.}}(k-k') \delta_{ss'} \;.
\ee
We consider matrix elements of the form
\be\label{starting-point-gravity}
    \Mbg\equiv \bra{\beta',\text{out}} T \ket{\beta,\text{in}} \;,
\ee
in which $T$ is now understood as the transition matrix in gravity, while $\beta$ and $\beta'$ are coherent states. The only real difference compared to gauge theory is that (looking ahead to the double copy) we consider asymptotic states with antisymmetric two-forms $B^{\mu\nu}$ (axions) and dilatons $\phi$, in addition to gravitons.
For this reason we introduce the label ``$P$" to indicate particle type \emph{and} quantum numbers. Explicitly, $P$ takes values in $ \{h^s, B^s, \phi \}$, and so runs over the polarisations of each type of particle. With this a coherent axio-dilaton gravity state can be written as
\be
    \ket{\beta} = e^{\sum_P \int_k \beta(k)_P a^\dagger_P(k)}\ket{0}e^{-\tfrac12 \sum_P \int_k |\beta_P(k)|^2}
\ee
in which $a_P$ and $\beta_P$ are the ladder operator and profile function for particle $P$.

{All steps now go through just as in gauge theory but with, essentially,} sums over $P$ appearing throughout. The expansion of $\Mbg$ in coherent state particles gives, much like~(\ref{gauge-amp-expanded}),
\begin{equation}\label{grav-amp-expanded}
\begin{split}
   \Mbgpert_{(n)} = \sum_r\frac{1}{r! (n - r)!}
   \int_{k_1 \dots k_n}   \sum_{\{ P_i \} } 
   \bar \beta'_{P_1}(k_1) \dots &\bar \beta_{P_r}(k_r)  \beta_{P_{r+1}}(k_{r+1}) \dots \beta_{P_n}(k_n)\times \\
   &\times \Mvac(\bar 1, \ldots, \bar r, r+1, \ldots n) \;.
\end{split}
\end{equation}
The story about disconnected and SMC diagrams is identical to what happens in gauge theory, and so in analogy with (\ref{SMC-first-def}) we have
\begin{equation}\label{SMC-gravity}
    \Mbg = e^W \sum_n \Mbgpert_{(n)}\Big|_{\text{SMC}} \;,
\end{equation}
in which, compare (\ref{W-definition}),
\be\label{W-definition-grav}
    W = -\frac12 \sum_P\int_k |\beta'_P(k)|^2 -\frac12 \sum_P |\beta_P(k)|^2
    +
    \sum_P \int_k {\bar \beta}'_P(k)\beta_P(k)\;.
\ee
The $\Mbgpert_{(n)}$ can, similarly to (\ref{gauge-amp-coherent-bgft}), be represented as amplitudes in a background,
\begin{equation}\label{grav-amp-coherent-bgft}
\begin{split}
    \Mbgpert_{(n)} = \frac{1}{n!} \int\! \deltahat^4\bigg( \Delta Q {- \sum_i k_i}\bigg)  \prod_j \dhat^4 k_i\,  \big[ h^{\mu_i \nu_i}(k_i) + &B^{\mu_i\nu_i}(k_i) + \phi^{\mu_i\nu_i}(k_i)  \big] \times \\
    & \times \Mstr_{\mu_1\nu_1 \dots \mu_n\nu_n} (-k_1, \dots, - k_n),
\end{split}
\end{equation}
in which the graviton, dilaton and B-field make up a `fat graviton'~\cite{Luna:2016hge}. 
(The dilaton polarisation vector is given by the flat metric plus gauge terms, see e.g.~\cite{Bern:2019prr}.)
The spin-stripped gravity amplitudes are again made by stripping polarisation vectors and momentum conserving delta functions from the vacuum amplitude. For $\Mvac$, the analogue of $\Avac$ in (\ref{gauge-amp-def-vacuum}), we have
\begin{equation}
    \Mvac({\bar 1}, \ldots, n) = {\bar\varepsilon}_{P_1}^{\,\mu_1 \nu_1} \dots \varepsilon_{P_n}^{\mu_n \nu_n} \deltahat^4\bigg( {\Delta Q + k_1 + \ldots - k_n}
    \bigg) \Mstr_{\mu_1 \nu_1 \dots \mu_n \nu_n}{(+k_1, \ldots, -k_n)} \;.
\end{equation}
The classical fields appearing are related to the coherent states as for gauge theory in (\ref{background-gauge-definition}). For gravitons for example,
\be\label{metric-pert-expect-val}
    h_{\mu\nu}(x) = \int_k \beta_s(k) \varepsilon_{\mu\nu}^s(k)e^{-ik\cdot x} + {\bar \beta}_s(k) {\bar\varepsilon}_{\mu\nu}^s(k)e^{ik\cdot x} \equiv \int\!\dhat^4k\, h_{\mu\nu}(k)e^{-ik\cdot x} \;,
\ee
which is just the metric perturbation (\ref{graviton-mode-expansion}) with $a_s(k)$ replaced by $\beta_s(k)$, and $a^\dagger_s(k)$ replaced by ${\bar \beta}^\prime_s(k)$. The metric $\eta_{\mu\nu}+\kappa h_{\mu\nu}$ satisfies the linearised Einstein equations.

At this stage there is no connection between the gauge theory coherent state profiles $\alpha$ and the gravity profiles $\beta$. This will change when we consider the double copy of our amplitudes.

%%%%%%%%%%%%%%%%%%%%%%%%%%%%
%%%%%%%%%%%%%%%%%%%%%%%%%%%%
\section{Double copy \label{sec:DC}}
%%%%%%%%%%%%%%%%%%%%%%%%%%%%
%%%%%%%%%%%%%%%%%%%%%%%%%%%%
%
%
While we do not intend to commit to any particular notion of vacuum double copy, it will be useful to introduce some key concepts, as well as provide a more explicit construction, in a concrete setting. For this reason we now briefly review BCJ double copy~\cite{Bern:2008qj,Bern:2010ue,Bern:2010yg,Bern:2022wqg,Bern:2019prr,Adamo:2022dcm}, highlighting those aspects which will be needed when we discuss the double copy of coherent states and background fields.

A tree-level gauge theory amplitude can be written as a sum over cubic diagrams,
\begin{equation}
    \Avac%^\text{tree}
    \sim \sum_i \frac{c_i n_i}{D_i},
\end{equation}
with $c_i$, $n_i$ colour and kinematic numerators, while $D_i$ are the corresponding propagators. The colour factors will obey some set of Jacobi relations. If it is possible to choose, e.g.~through some generalised gauge transformation, the $n_i$ such that they satisfy {those} same relations, the amplitude has colour-kinematic duality. Such a representation always exists for tree-level amplitudes in pure Yang-Mills, for instance. Replacing the colour factors in such amplitudes with a new set of kinematic numerators, $c_i \to \tilde n_i$, results in an amplitude belonging to some gravitational theory.

This operation is quite flexible, and the $\tilde n_i$ can differ from the $n_i$ in various ways -- they can come from the same amplitude written in a different generalised gauge, correspond to different external states, or belong to a different gauge theory altogether. The form of the $n_i$ and choice of $\tilde n_i$ decides the precise axio-dilaton gravity theory, or \textit{target} theory, that the resulting amplitude belongs to. For examples, see the lists of related theories in~\cite{Bern:2019prr}.

\subsection{Polarisation and colour}\label{sec:DC-chat}

Having a more explicit picture of how this double copy map handles polarisation and colour of external gluons will be useful below, when we return to coherent states.
To begin, consider an amplitude $\Avac(1_a^s)$ of some collection of matter particles and, for now, one incoming gluon of colour $a$ and polarisation $s$.
The colour factor for each cubic diagram is some product of generators $\sfT$ in appropriate representations. {To highlight the dependence of the colour factors on the external gluon colour $a$ we write $c_i\equiv c_i^a$.}
{Note that whatever Jacobi identity  holds does so regardless of the explicit value of $a$.}
Similarly, the kinematic numerator depends on the gluon polarisation, but only through $\varepsilon_\mu^s$. We again decorate it accordingly as $n_i^s$, and introduce a spin-stripped kinematic numerator $n^s_i \equiv \varepsilon_\mu^s n_i^\mu$, where $n_i^\mu$ contains all of the spin-independent kinematics.
(This is always possible, even for the diagrams with disconnected propagators~(\ref{propagator-gauge}) which do not explicitly contain the polarisation vectors. They can be exposed through the normalisation condition $ \varepsilon^s_\mu  \bar \varepsilon_{s'}^\mu = - \delta^s_{s'}$, and $\varepsilon_{\mu\nu}^P \bar\varepsilon_{P'}^{\mu\nu} = \delta^P_{P'}$ for the equivalent diagrams in gravity.)

Equipped with this, we can write the amplitude as
\begin{equation}
    \Avac%^\text{tree}
    (1_a^s)  
    \sim \sum_i \frac{c_i^a \varepsilon^s_\mu n_i^\mu}{D_i}.
\end{equation}
If the $n_i$ exhibit colour-kinematic duality, as we assume throughout, the standard double copy replacement $c_i^a \to \tilde n_i^s \equiv \tilde \varepsilon_\mu^s \tilde n_i^\mu$ turns the gauge theory amplitudes into gravitational amplitudes. Explicitly,
\begin{equation}\label{output-of-DC}
    \Avac(1_a^s) 
    \DCto \sum_i \frac{ \varepsilon^s_\mu \tilde \varepsilon_\nu^s    n_i^\mu \tilde n_i^\nu}{D_i}.
\end{equation}
The first point to highlight is that the colour factor is replaced in the same way, independent of $a$, hence the output of the double copy is identical regardless of gluon colour. Second, in removing the colour index we instead introduced a second $s$; this is not, note, summed over, nor does it directly label particle spin or helicity. {Its precise role should become clear from the general construction and examples immediately below.}

The external states $\tilde \varepsilon_\mu^s$ need not be equal to the original polarisation vectors. Nevertheless, we can decompose the product $\varepsilon_\mu^s {\tilde \varepsilon_\nu^s}$ into symmetric traceless, anti-symmetric and trace parts, corresponding to external gravitons, $B$-fields and dilatons respectively. Further decomposing each of these parts into a basis of polarisation tensors, we write
\begin{equation}\label{BCJ-pol-decomp}
    \varepsilon_\mu^s(k) \tilde \varepsilon_\nu^s(k) = \sum_P \C^s_P \varepsilon_{\mu\nu}^P(k),
\end{equation}
for some constants $\C_P^s$ which depend on the double copy scheme through the choice of $\tilde \varepsilon_\mu^s$ and retain the spin index of the original polarisation vector. Applying this to the output of the double copy~(\ref{output-of-DC}), one obtains a linear combination of gravity amplitudes,
\begin{equation}\label{1-point-DC}
\begin{split}
     \Avac%^\text{tree}
     \left(1^s_a\right) \DCto  \sum_P \C^s_P \Mvac%^\text{tree}
     (1_P) \;.
    \end{split}
\end{equation}
Some simple examples make this explicit. First, suppose the gluon has positive helicity, $\varepsilon^{+}_\mu$. If we choose kinematic numerators such that $\tilde \varepsilon_\mu^{+} = \varepsilon_\mu^{+}$, then since the product of positive-helicity gluon polarisations is equal to the positive-helicity graviton polarisation tensor, $\varepsilon_\mu^{+} \varepsilon_\nu^{+} = \varepsilon^{h^+}_{\mu\nu}$, the only non-zero constant is $\C_{h^+}^+ = 1$. Hence an amplitude with a positive helicity gluon directly double copies to that of a positive helicity graviton,
\begin{equation}\label{BCJ-1-particle-amp-DC}
    \begin{split}
        \Avac%^\text{tree}
        \left(1_a^+\right) \DCto  \Mvac%
        %^\text{tree}
        \left(1_h^+\right).
    \end{split}
\end{equation}
If we instead choose a linearly polarised gluon with a symmetric double copy scheme, the product of polarisation vectors is not traceless; calling the two linear polarisations $s=1,2$, we separate out the trace part and find
\begin{equation}
\begin{split}
        \varepsilon_\mu^1 \varepsilon_\nu^1 &=  \frac{\varepsilon^1_\mu \varepsilon^1_\nu - \varepsilon^2_\mu \varepsilon^2_\nu}{2} + \frac{\varepsilon^1_\mu \varepsilon^1_\nu + \varepsilon^2_\mu \varepsilon^2_\nu}{2} = 
        \frac{
        {\left( \varepsilon_h^\oplus \right)_{\mu\nu}}
        + \left( \varepsilon_\phi  \right)_{\mu\nu}}{\sqrt2}  \;,
\end{split}
\end{equation}
the first term corresponding to a ``plus''-polarised graviton and the second to a dilaton. Now there are two non-zero constants, ${\C_{h^\oplus}^1} = 1/\sqrt2$ and $\C_\phi^1 = 1/\sqrt2$, and so the gauge theory amplitude double copies to
\begin{equation}
    \Avac\left(1^1_a\right) \DCto \frac{1}{\sqrt2} \Mvac\left({1_h^\oplus} \right) + \frac{1}{\sqrt2} \Mvac\left(1_\phi\right).
\end{equation}

%%%%%%%%%%%%%%%%%%%%%%%%%%%%%%%%%%%%%%%%%%%%%%%%%%%%%%%%%%%%%%

\subsection{Double copy for backgrounds}
We turn now to the double copy of amplitudes on those backgrounds that can be represented as asymptotic coherent states, the goal being to construct a map $\Abg \DCto \Mbg$ from (\ref{starting-point}) to (\ref{starting-point-gravity}). It seems clear from the coherent state picture that if such a double copy map exists then it {should} be largely inherited from the double copy on flat space. Our approach will be to maximally utilise vacuum double copy relations, whilst introducing a minimum of  additional structure: we simply double copy all of the flat space amplitudes in~(\ref{Abg-from-Astr}), made explicit in the expansion~(\ref{gauge-amp-expanded}).
This approach requires consistently double copying an \emph{infinite} number of vacuum amplitudes of different multiplicities, inside a sum, rather than double copying a single chosen amplitude. 

In order to do so, we introduce two restrictions on the vacuum double copy map:
\begin{enumerate}
    \item The double-copied vacuum amplitudes should all belong to the \emph{same} target theory.
    \item Polarisation vectors should be consistently chosen, i.e.~new polarisation vectors $\tilde \varepsilon_\mu^{s_1}$ and $\tilde \varepsilon_\mu^{s_2}$ are equal if polarisations $s_1$ and $s_2$ are equal. For outgoing gluons we choose, naturally, the complex conjugates of the $\tilde{\varepsilon}^s_\mu$.
\end{enumerate}
In the context of the BCJ double copy, both constraints regard the kinematic numerators $n_i$ and $\tilde n_i$. The first deals with the spin-stripped kinematic numerators which must be chosen, across different amplitudes, so that the object we have called $\Mvac$ really is a vacuum amplitude in a given target theory. The second constraint demands consistency of how external states are double copied.
Together they ensure that the extension of~(\ref{1-point-DC}) to higher multiplicity and outgoing gluons is
\begin{equation}\label{n-point-DC}
    \Avac(\bar 1, \ldots , n) \DCto \sum_{P_1 \dots P_n} \bar \C_{P_1}^{s_1} \dots \C_{P_n}^{s_n} \Mvac(\bar 1, \dots, n),
\end{equation}
in which the same set $C_P^s$ appears consistently for all amplitudes.

{Henceforth, we take~(\ref{n-point-DC}) to be the defining feature of vacuum double copy; it is just some map which transforms gauge theory amplitudes into amplitudes in a gravitational theory, with a weighted sum over external states. It also means we can be agnostic about its precise mechanics. While the BCJ double copy is traditionally limited to tree-level amplitudes and axio-dilaton gravity, there is no problem in using extensions in which dilaton and $B$-field propagators are removed~\cite{Johansson:2014zca,Luna:2017dtq,Bern:2019nnu}, or BCJ at higher loop order provided the BCJ conjecture holds~\cite{Bern:2010ue}. We could also use any other scheme relating vacuum amplitudes through~(\ref{n-point-DC}), and our map directly inherits the range of applicability of that scheme.}

We are now in a position to explore the double copy of the gauge theory amplitudes $\Abg_{(n)}$ and their connection to the gravity amplitudes $\Mbg_{(n)}$. We will see that there is a natural double copy map between them. Beginning at leading order, we apply the double copy to all the vacuum amplitudes appearing in $\Abg_{(1)}$, accounting for the restrictions above, which yields
\begin{equation}\label{gauge-bg-DC-LO}
\begin{split}
    \Abgpert_{(1)} =     & \sum_{a,s}  \int_{k}  \bar \alpha^{\prime a}_{s} \Avac(\bar1) + \alpha^a_{s} \Avac(1)
    \quad 
    \DCto \quad \sum_{a,s,P}
    \int_k  \bar \alpha^{\prime a}_{s} \bar \C_P^s \Mvac(\bar1_P) + \alpha_{s}^a  \C_P^s  \Mvac(1_P) \;.
\end{split}
\end{equation}
This is to be compared to the $n=1$ gravity amplitude (\ref{grav-amp-coherent-bgft}),
\begin{equation}\label{Mgbpert-1}
    \Mbgpert_{(1)} = \sum_P  \int_k \bar \beta'_P \Mvac(\bar1_P) + \beta_P \Mvac(1_P) \;,
\end{equation}
which has precisely the same structure in terms of amplitudes.
In fact the two expressions (\ref{gauge-bg-DC-LO}) and  (\ref{Mgbpert-1}) are precisely equal provided that the gravitational and gauge theory coherent state profiles are related by
\begin{equation}\label{profiles-DC-key}
    \beta_P(k) = \sum_{a,s} C_P^s  \alpha^a_s(k) \;, \qquad {\beta'_P(k) = \sum_{a,s} C_P^s  \alpha^{\prime a}_s(k)} \;. 
\end{equation}
The structure and consequences of this result will be discussed shortly, but first we note that, going to higher $n$, we can use~(\ref{n-point-DC}) to replace all the vacuum amplitudes in~(\ref{gauge-amp-expanded}), {resulting in the double copy
\begin{equation}\label{theexpression}
\begin{split}
    \Abgpert_{(n)} &\DCto \sum_r \frac{1}{r! (n - r)!}  \int_{k_1 \dots k_n} {\sum_{\{a_i,s_j,P_k\}}}
    \bar C_{P_1}^{s_1} \bar \alpha_{s_1}^{\prime a_1}(k_1) \dots  C_{P_n}^{s_n} \alpha_{s_n}^{a_n} (k_n) \Mvac(\bar 1\ldots\bar r, r+1 \ldots n) \;,
\end{split}
\end{equation}
The sums over spin and colour may then be performed using the leading-order relation~(\ref{profiles-DC-key}) between profile functions, rendering (\ref{theexpression}) equal to
\begin{equation}
    \begin{split}
\Abgpert_{(n)} &\DCto \sum_r \frac{1}{r! (n - r)!}  \int_{k_1 \dots k_n} {\sum_{\{P_i\}}}
    \bar \beta_{P_1}(k_1) \dots  \beta_{P_n}(k_n) \Mvac(\bar 1\ldots\bar r, r+1 \ldots n)  \;,
\end{split}
\end{equation}
which is identical to~(\ref{grav-amp-expanded}).} Consequently $\Abgpert_{(n)} \to \Mbgpert_{(n)}$ under our double copy map.
The effect on the full gauge theory amplitude~(\ref{Abg-from-Astr}) of double copying all vacuum amplitudes is
\begin{equation}\label{315}
    \Abg \DCto e^{ -\tfrac12  \int_k |\alpha^a_s|^2 + |\alpha^{\prime a}_{s}|^2    } \sum_n \Mbgpert_{(n)},
\end{equation}
which is equal to the gravitational coherent-state amplitude~(\ref{SMC-gravity}) up to the leading exponential constant -- effectively, (\ref{315}) represents a matrix element where the gravitational coherent states are not correctly normalised.   
%
%which would be equal to the gravitational coherent-state amplitude~(\ref{starting-point-gravity}) were it properly normalised.

Up to this normalisation (which we will address below), we see that vacuum double copy has a natural extension in which coherent state amplitudes in gauge theory double copy to coherent state amplitudes in gravity. In particular, the coherent state of gluons $\ket{\alpha}$ is mapped to (again, but only for now, up to normalisation) the gravitational coherent state $\ket{\beta}$ through (\ref{profiles-DC-key}). We emphasise that this was not an input to the calculation -- we did not insist that coherent states should map to coherent states.

The sum over colour appearing in (\ref{profiles-DC-key}) is curious. It stems directly from the fact that gluons of each colour are all copied identically and independently, recall Sec.~\ref{sec:DC-chat}. As a result, the same coherent state $\ket{\alpha}$ may result in different explicit expressions for~(\ref{gauge-bg-DC-LO}) and~(\ref{profiles-DC-key}) depending on the chosen colour basis -- as odd as this sounds, the same is true of the choice of polarisation basis. This could also be seen as a positive, though, as it adds to the flexibility of the scheme.
We will discuss this again, along with other possible choices, in Sec.~\ref{sec:sq-or-not}.

Returning to the normalisation factor, we could argue that this should be fixed by hand -- it comes after all from the asymptotic coherent states, not from any amplitude, so there is no expectation it should obey a double copy relation. There is, however, a more elegant method, and one which does not require assuming a perturbative expansion. 
 Returning to our starting point~(\ref{starting-point}) and (\ref{starting-point-gravity}), we observe that the coherent state amplitudes
\begin{equation}\label{amp-DC-ladder-T}
\begin{split}
    \Abg &= \bra{\text{out}}   \exp\bigg( \int_k  a_s^a \bar \alpha_s^{\prime a} - a_s^{ a\dagger } \alpha_s^{\prime a} \bigg)   T_\text{gauge}  \exp\bigg( \int_k a_s^{a\dagger } \alpha_s^a - a_s^a \bar \alpha_s^a \bigg)   \ket{\text{in}}, \\
    \text{and} \\
    \Mbg &= \bra{\text{out}}  \exp\bigg( \int_k  a_P \bar \beta'_P - a_P^\dagger \beta'_P \bigg) T_\text{gravity}  \exp\bigg( \int_k a_P^\dagger \beta_P - a_P \bar \beta_P \bigg)  \ket{\text{in}},
\end{split}
\end{equation}
enjoy a clear double copy relation $\Abg\to \Mbg$ through the replacement {rules}
\begin{equation}\label{replacement-ladder-T}
T_\text{gauge} \to T_\text{gravity} \;, \qquad    a_s^{a\dagger} \to \sum_P C_P^s a^\dagger_P \;.
\end{equation}
The first of these is essentially the vacuum double copy, originating, in the BCJ context, from the choice of spin-stripped kinematic numerators. This is %clearly
separated from the second rule, stemming from the choice of new polarisation vectors $\tilde \varepsilon^s_\mu$, which defines a classical double copy relation transforming the asymptotic gauge field into asymptotic axions, dilatons and gravitons. Together these rules reproduce all the perturbative results above and automatically generate the correct normalisation.

In light of the replacement rules~(\ref{replacement-ladder-T}) the double copy map naturally extends to cases where the external states $\ket{\text{in}}$ and $\ket{\text{out}}$ contain number states of gluons/gravitons. We simply replace the creation operator of each such external gluon in a similar manner. The doubled polarisation vectors $\tilde \varepsilon_\mu$ of these external gluon must be chosen in the same way across all amplitudes, though this choice can be decoupled from that of the new polarisation vectors for gluons in the  coherent states.

Though~(\ref{replacement-ladder-T}) are perhaps conceptually the clearest replacement rules, it is convenient for practical purposes to consider how they relate amplitudes in the background field picture~(\ref{gauge-amp-coherent-bgft}) and~(\ref{grav-amp-coherent-bgft}).
The replacement of the $T$-matrix in (\ref{replacement-ladder-T}), realised by some vacuum double copy scheme, relates spin-stripped amplitudes in the two theories,
\begin{equation}
    \Astr_{\mu_1 \dots \mu_n}^{a_1 \dots a_n} (k_1, \ldots, k_n) \DCto \Mstr_{\mu_1\nu_1 \dots \mu_n \nu_n}(k_1, \ldots, k_n) \;,
\end{equation}
while the ladder operator replacement in (\ref{replacement-ladder-T}) results in the gauge theory background (\ref{background-gauge-definition}) being mapped to the fat graviton
\begin{equation}\label{classic-DC-field}
\begin{split}
    A_\mu(x) &\DCto
   \sum_P \int_k
    \beta_P(k)\varepsilon_{\mu\nu}^P(k) e^{-ik\cdot x}
    +
    {\bar\beta}_P(k) {\bar \varepsilon}_{\mu\nu}^P(k) e^{+ik\cdot x} \\
&=  h_{\mu\nu}(x) + B_{\mu\nu}(x) + \phi_{\mu\nu}(x) \;,
\end{split}
\end{equation}
with $\beta_P$ defined through~(\ref{profiles-DC-key}). We turn now to a discussion of this relation.

%%%%%%%%%%%%%%%%%%%%%%%%%%%
\section{Emergence of classical double copy}\label{sec:classicalDC}
%%%%%%%%%%%%%
\subsection{To square or not to square}\label{sec:sq-or-not}
%%%%%%%%%%%%%%
%%%
%
Our scheme for double copying coherent state amplitudes naturally leads to~(\ref{classic-DC-field}), which we may interpret as defining a classical double copy arising from an amplitudes-based approach.  Our map says that for any background gauge field with a coherent state representation, we should remove the group generators and double the polarisation vectors to find a classical double copy in gravity. Scattering amplitudes on these gauge and gravity backgrounds are then related through the double copy map described above.

Applying~(\ref{classic-DC-field}) to specific examples, below, we will see that we recover some known results from the literature. However, there are other possible definitions of classical double copy, and many literature examples do \emph{not} follow the structure found here. One could consider, motivated by e.g.~the peculiar colour-basis dependence in (\ref{profiles-DC-key})  and the example in~\cite{Monteiro:2020plf}, whether the profile functions $\alpha$ should also be doubled, so that the gravitational background
\be
\beta \overset{?}{\sim} \alpha^a \alpha^a
\ee
is formed from a colour invariant. Such constructions require the introduction of some extra structure (not least to achieve the correct mass dimension). This often comes in the form of a biadjoint scalar~\cite{Anastasiou:2014qba,Ilderton:2024oly}, though it is not obvious from the outset how it should be chosen.

Another option is to double only part of the profile functions. That is the approach taken in~\cite{Monteiro:2020plf}, where a classical double copy between Coulomb and Schwarzschild in split signature is studied. Though neither of these fields have a coherent state description ({they are sourced}, and therefore do not fit into our framework), their late-time behaviour does, with profile functions
given by three-point amplitudes in vacuum, $\alpha \sim \Avac_3$ for Coulomb and $\beta \sim \Mvac_3$ for Schwarzschild~\cite{Monteiro:2020plf,Adamo:2022ooq}. The two are related through {the double copy} of the three-point \emph{amplitudes} appearing in the profile functions, very much resonating with the philosophy of our approach.

The distinction between that case and ours is that $N$-point amplitudes on Coulomb/ Schwarzchild are approximations to $N+2$-point \emph{vacuum} amplitudes in the leading self-force expansion where one massive particle is very much heavier than the other. (For double copy in this context see e.g.~\cite{Haddad:2020tvs,Brandhuber:2021kpo, Brandhuber:2021eyq}.) {In contrast}, the coherent states we have considered are built from asymptotically free gluons and gravitons rather than being generated by a point source, and the coherent state profile functions play a role more akin to wavepackets, which are not typically doubled. 

The question of whether or not to square also arises in the Kerr-Schild double copy, as the division of some gauge field $A_\mu =  v_\mu \phi$ into a vector (which is doubled) and a scalar (which is not) is inherently ambiguous. For certain classes of fields our approach settles this ambiguity by saying that, in momentum space, only polarisation vectors should be copied. We will see this as part of the examples below. For another approach see~\cite{Kent:2025pvu}, with which it would be interesting to compare.

%%%%%%%%
\subsection{Plane waves}
%%%%%%%%%
The simplest example illustrating that some known classical double copy relations emerge from our coherent state picture is that of plane waves~\cite{Monteiro:2014cda}.
In gauge theory these are exact vacuum solutions of the Yang-Mills equations and are functionally abelian, being valued in the Cartan of the gauge group~\cite{Adamo:2017nia}. 
On the gravity side, plane waves arise not just as models of gravitational waves but approximate any curved spacetime in the neighbourhood of a null geodesic via the Penrose limit~\cite{Penrose}, the single copy of which has been explored in~\cite{Chawla:2024mse, Chawla:2025uwu}.

When discussing plane waves it is convenient to work in lightfront coordinates,
\be
    \ud s^2 = 2 \ud u \ud v - \ud x^i \ud x^i \;,
\ee
where $i \in \{1,2\}$. Gauge theory plane waves may be represented by `Rosen gauge' potentials of the form, c.f.~(\ref{background-gauge-definition}),
\begin{equation}\label{plane-wave-potential}
    A_\mu(x) = \sfT \delta_\mu^i a_i(u),
\end{equation}
in which $a_i(u)$ are two real functions of `lightfront time' $u$ and $\sfT$ is a Cartan generator. (There is no added complication in generalising to a sum over such generators.)
The corresponding chromo-electromagnetic fields are null, and polarised transverse to the direction of propagation of the wave.
The potential (\ref{plane-wave-potential}) can be written as a Fourier integral over on-shell modes, for which it is convenient to work in a basis of lightfront helicity vectors
\be\label{def:helicuty-pol}
    \varepsilon^s_\mu = \frac{1}{\sqrt{2}} \bigg( \frac{k_1 + i s k_2}{k_u} , 0, 1, i s \bigg) \;, \qquad s=\pm 1 \;,
\ee
with vector components in order $\mu=(u, v, 1, 2)$.
With this, and defining helicity-basis profile functions $a_s(u)= {(a_1(u) - i s a_2(u))/\sqrt{2}}$, the potential (\ref{plane-wave-potential}) {becomes}
\begin{equation}\label{plane-wave-on-shell}
    A_\mu(x) = \sfT\int_k  {\deltahat(k_1)\deltahat(k_2)}\, (2k_u) \varepsilon^s_\mu\, a_s(k_u)\,   e^{-ik\cdot x} + \cc \;,
\end{equation}
in which the on-shell condition (in lightfront coordinates) is $k_v = {k_i k_i}/(2k_u)$ and the profile functions have Fourier transform
\begin{equation}
 a_s(u) = {\int_{-\infty}^\infty {\hat \ud} k_u} \, a_s(k_u) e^{-ik_u u} \;.
\end{equation}
Note that the helicity vectors are constant, as $\varepsilon_\mu^s(k) \to (0,0,1, is)/\sqrt{2}$ on the support of the delta functions in (\ref{plane-wave-on-shell}).

Choosing a symmetric double copy scheme, we make no changes to the integrand of (\ref{plane-wave-on-shell}) except to double the polarisation vectors as $\varepsilon^{+}_\mu \to \varepsilon^{+}_\mu \varepsilon^{+}_\nu = \varepsilon_{\mu\nu}^{+} $ and $\varepsilon^{-}_\mu \to \varepsilon^{-}_\mu \varepsilon^{-}_\nu = \varepsilon_{\mu\nu}^{-}$, both automatically symmetric and traceless such that only a graviton field is generated. As the polarisation vectors in this case are effectively constant, the Fourier integrals are the same as in the gauge case and we find the metric perturbation
\begin{equation}
    h_{\mu\nu}(x) =  \varepsilon_{\mu\nu}^s a_s (u) \;.
\end{equation}
This results in the metric
\begin{equation}
\label{rosen-gauge-metric}
   \ud s^2 = 2 \ud u \ud v - \gamma_{ij}(u) \ud x^i \ud x^j,
\end{equation}
where
\begin{equation}
\label{plane-wave-Rosen}
    \gamma_{ij} = \delta_{ij} - \frac{\kappa}{\sqrt{2}}
    \begin{bmatrix}
        a_1(u)  & a_2(u) \\
        a_2(u) & -a_1(u)
    \end{bmatrix} \;,
\end{equation}
which is indeed a plane wave, though not necessarily a vacuum solution, recovering the results of~\cite{Monteiro:2014cda}.
The extra factor of $\sqrt{2}$ compared to the gauge potential comes, in effect, from insisting that the doubled polarisation vectors are correctly normalised.

This is not, of course, the only possible (quantum and classical) double copy scheme for plane wave backgrounds~\cite{Adamo:2017nia,Chawla:2024mse,Chawla:2025uwu}. A prescription has been found which double copies gauge theory amplitudes on Yang-Mills plane waves to gravitational amplitudes on gravitational plane waves~\cite{Adamo:2017nia}, thus relating vacuum solutions on both sides of the correspondence. {It} has been verified up to three points~\cite{Adamo:2017nia,Adamo:2020qru}, with non-trivial memory effects appearing already at two points~\cite{Zhang:2017rno,Adamo:2022rmp,Cristofoli:2025esy,Klisch:2025mxn}, and a generalisation of colour-kinematic duality to plane wave backgrounds has been studied in~\cite{Adamo:2018mpq}. This double copy is typically discussed in Brinkmann gauge where, instead of (\ref{plane-wave-potential}), we use the gauge-equivalent potential
    \be\label{Brinkmann-gauge}
        A_\mu(x) = - \sfT n_\mu  x^i a'_i(u)\;.
    \ee
On the gravitational side the metric is (abusing notation, since these are not the same coordinates as in (\ref{rosen-gauge-metric}))
\be\label{Brinkmann-gravity}
\ud s^2 = 2 \ud u \ud v - \ud x^i \ud x^i - H_{ij}(u)x^i x^j\ud u^2 \;,
\ee
where $H_{ij}$ is symmetric and traceless but otherwise arbitrary. In particular, it is not constrained to be related to the potential (\ref{Brinkmann-gauge}). 
Hence it is unclear how this double copy scheme relates to ours. For instance, it is not obvious how to write (\ref{Brinkmann-gauge}) or (\ref{Brinkmann-gravity}) in terms of on-shell polarisations as their Fourier profiles, which contain derivatives of delta functions, must be regulated. However, doing so typically takes us out of the space of plane waves.

Ultimately {we learn} that there are multiple ways to define double copy on plane wave backgrounds, and it may be enlightening to explore the connections between them. In this regard we note that the gauge transformation relating Rosen and Brinkmann is \emph{large}, and perturbative coherent state methods seem to prefer the former gauge choice~\cite{Cristofoli:2022phh}. It may be, then, that connecting these different notions of double copy would shed some light on relations between gauge and diffeomorphism invariance~\cite{Anastasiou:2014qba,Anastasiou:2018rdx}. For now, though, we turn to other examples.
%

%%%%%%%%%%%%%%%%%%%%%%%%%%%%%%%%%%%%%%%%%%%%%%%%%%%%
\subsection{Self-dual backgrounds and Kerr-Schild}\label{sec:kerr-schild}
%%%%%%%%%%%%%%%%%%%%%%%%%%%%%%%%%%%%%%%%%%%%%%%%%%%%
%
We consider next coherent state profiles $\alpha$ and $\alpha'$ chosen such that the corresponding background (\ref{background-gauge-definition}) is \emph{self-dual}. To do so we use a helicity basis, and choose the profiles such that we only have incoming positive-/outgoing negative-helicity gluons, hence
\begin{equation}\label{A-is-self-dual}
    A_\mu(x) = \sfT \int_k \varepsilon^{+}_\mu {\bar \alpha}^{\prime}(k) \, e^{i k \cdot x} + \varepsilon^{+}_\mu  \alpha(k) \, e^{- i k \cdot x}\;,
\end{equation}
in which only the positive helicity polarisation vector appears, while $\alpha^{\prime}(k)$ and $\alpha(k)$ are arbitrary. Note that $A_\mu$ is complex in general. We will again assume the field is in the Cartan, or in a single colour direction, as the colour plays little role in what follows. $A_\mu$ is then an exact and \emph{self-dual} vacuum solution. (See ~\cite{Alfonsi:2020lub} for non-abelian classical double copy examples, and~\cite{Armstrong-Williams:2024icu} for motivation to study abelian sectors.)

To relate this to our notion of double copy and to the literature on self-dual double copy, we follow~\cite{Tod:1982mmp}. Let $\phi(x)$ be a solution of the massless wave equation in Minkowski space, $\partial^2\phi=0$. Acting on $\phi$ with the `spin raising' operator
\be
	\hat{O}_\mu = \frac{1}{\sqrt{2}} \bigg({\pd_1 + i \pd_2} , 0, {\pd_u}, i{\pd_u}  \bigg)  \;,
\ee
generates a self-dual solution $A_\mu = \hat{O}_\mu \phi$ of the Maxwell equations. Suppose that, in addition, $\phi(x)$ obeys
\be\label{Plebanski}
	\partial_u^2\phi \partial_z^2\phi = (\partial_u\partial_z\phi)^2 \;,
\ee
where $z=(x^1+i x^2)/\sqrt{2}$. This is the Plebanski equation of self-dual gravity~\cite{Plebanski:1975wn}. It is then automatic that
\be\label{self-dual-fields-0}
	   g_{\mu\nu} = \eta_{\mu\nu} + \kappa {\hat O}_\mu \hat{O}_\nu \phi(x) \;,
\ee
is a self-dual solution of the Einstein equations~\cite{Tod:1982mmp,Monteiro:2014cda,Berman:2018hwd,Brown:2023zxm}.

Now, the choice of spin-raising operator is not unique. For us it is more convenient to instead consider the operator
\be\label{KSoperator}
	\hat{V}_\mu = \frac{1}{\sqrt{2}} \bigg(\frac{\pd_1 + i \pd_2}{\pd_u} , 0, 1,i  \bigg) \;.
\ee
$\hat{V}_\mu$ is evidently non-local in position space (a point we return to below), but its action on any object is clear if we go to Fourier space; it inserts exactly the positive-helicity polarisation vector~(\ref{def:helicuty-pol}) under the Fourier integral. It follows immediately that, given a scalar field $\varphi(x)$, the metric
\be\label{self-dual-fields-1}
	g_{\mu\nu} = \eta_{\mu\nu} + \kappa {\hat V}_\mu \hat{V}_\nu {\varphi(x)} \;,
\ee
is a self-dual solution of the Einstein equations if $\varphi(x)$ solves the wave equation,
\be\label{KS-scalar}
    \varphi(x)=\int_k {\bar \alpha}^{\prime}(k) \, e^{i k \cdot x} + \alpha(k) \, e^{- i k \cdot x} \;,
\ee
and obeys the analogue of (\ref{Plebanski}) resulting from the different derivative factors in $\hat{O}_\mu$ and $\hat{V}_\mu$, that is $\varphi \partial_u^{-2}\partial_z^2\varphi = ( \pd_u^{-1} \partial_z\varphi)^2$. The corresponding single copy of the metric, exchanging one factor of ${\hat V}_\mu$ for a generator $\sfT$, is precisely the self-dual background (\ref{A-is-self-dual}),
\be\label{single-copy}
    A_\mu(x)
    = \sfT\,\hat{V}_\mu \int_k {\bar \alpha}^{\prime}(k) \, e^{i k \cdot x} + \alpha(k) \, e^{- i k \cdot x}
    = \sfT\,{\hat V}_\mu \varphi(x)   \;.
\ee
Note that with this construction the scalar, gauge field and graviton have the correct mass dimension.

On the one hand, the gauge field (\ref{single-copy}) and metric (\ref{self-dual-fields-1}) are related by the `operator' Kerr-Schild double copy~\cite{Monteiro:2014cda}: $\hat V$ obeys
\be
    [ {\hat V}_\mu , {\hat V}_\nu] = 0 \;, \qquad
    \eta_{\mu\nu} \hat{V}_\mu \hat{V}_\nu = 0 \;, \qquad
    \hat{V}^\mu\partial_\mu = 0 \;,
\ee
as required. On the other hand, trading any generators $\sfT$ in $A_\mu$ for a copy of $\hat V_\nu$ is equivalent to doubling the polarisation vector $\varepsilon^{+}_\mu \to \varepsilon^{+}_{\mu\nu}$ under the Fourier integral, 
\begin{equation}
    {\hat V}_\mu \hat{V}_\nu {\varphi(x)} =   \int_k \bar\alpha'(k) \varepsilon^{+}_{\mu\nu} e^{i k \cdot x} + \alpha(k) \varepsilon^{+}_{\mu\nu} e^{-i k \cdot x} \;,
\end{equation}
which is exactly our prescription for the classical double copy -- the polarisation vectors are doubled, the profile functions are not. This makes explicit the overlap between our notion of classical double copy and the literature on self-dual, Kerr-Schild double copy.

\medskip
\noindent For an arbitrary self-dual gauge field, our double copy construction generates a linearised solution to the Einstein equations. This is to be expected, as the metric is built from free, on-shell gravitons. However, for the class of gauge fields additionally obeying the Plebanski equation, our amplitudes-based classical double copy between~(\ref{single-copy}) and~(\ref{self-dual-fields-1}) constitutes a map between exact vacuum solutions in both gauge theory and gravity, as well as any-multiplicity scattering amplitudes on those backgrounds.

%%%%%%%%%%%%%%%%%%%%
\subsubsection{Example: focus waves}
We can give a non-trivial example of the above construction in which all integrals can be performed explicitly. We take $\alpha'=0$ and
\be
    \alpha(k) = f(k_u) \frac{4 \pi}{\kk } 
	 \exp \left(-\frac{k_\LCperp k_\LCperp}{2 \kk k_u}\right) \;,
\ee
in which $\kk$ is a positive constant and $f$ is an arbitrary function. This defines a `focus wave' solution of the massless wave equation, i.e.~$\partial^2\varphi(x)=0$ where~\cite{Hillion92}:
\be\label{FF-scalar}
\begin{split}
    \varphi(x) &=  \int_k e^{-ik\cdot x} \alpha(k) \equiv \int\!\frac{\ud^2k_\LCperp}{(2\pi)^2}\int\limits_0^\infty\!\frac{\ud k_u}{(2\pi)2k_u} e^{-ik\cdot x} \alpha(k) \\
	  &= \frac{1}{1+i \kk v} \int\limits_0^\infty\!\frac{\ud k_u}{(2\pi)} f(k_u)e^{-i k_u \big(u - \frac{\kk}{2}\frac{x^\LCperp x^\LCperp}{1+i \kk v}\big)}\\
      &=  
    \frac{1}{1+i \kk v}f\bigg(u - \frac{\kk}{2}\frac{x^\LCperp x^\LCperp}{1+i \kk v}\bigg) \;.
\end{split}
\ee
A well-studied example is the choice $f(x) = \exp(-i\omega x)$; promoted to a solution of Maxwell's equations {(see below)}, $\varphi$ is an example of a `flying focus' wave with spatial Gaussian focussing, in which the group and phase velocities of the wave are in opposite directions (moving at the speed of light). See~\cite{Adamo:2025vzv} for a recent discussion of self-dual focus waves, their application and further references.

The corresponding self-dual gauge field is
\be\label{FF-gauge}
	A_\mu(x) = {\sfT}\int_k \varepsilon^{+}_\mu(k) \varphi(k) =   {\sfT}\mathcal{E}_\mu(x) \varphi(x) \;,
\ee
in which
\be
	\mathcal{E}_\mu(x) = \frac{1}{\sqrt{2}}\bigg(-i\frac{\kk (x^1+ i x^2)}{ 1+i \kk v},0,1,i\bigg) \;.
\ee
It is easily checked that $\varphi(x)$ obeys {the Plebanski equation}, so the classical double copy of $A_\mu$ which arises from the coherent state picture is a self-dual solution of the Einstein equations: we find explicitly
\be\begin{split}\label{FF-gravity}
	g_{\mu\nu} &= \eta_{\mu\nu} + \kappa 
\int_k \varepsilon^{+}_\mu(k)\varepsilon^{+}_\nu(k) \varphi(k) \\
    &= \eta_{\mu\nu} + \kappa \, \mathcal{E}_\mu(x)\mathcal{E}_\nu(x) \varphi(x) \;,
\end{split}
\ee
where the Fourier integral can either be computed explicitly or sidestepped by realising that $[{\hat V}_\mu,\mathcal{E}_\nu]=0$ and that, from (\ref{FF-gauge}), our $\varphi(x)$ is an \emph{eigenvector} of ${\hat V}_\mu$.  Thus the considered self-dual fields obey `multiplicative' Kerr-Schild double copy in momentum space \emph{and} position space -- simpler than the convolutional structure which might be expected from locality of amplitudes in momentum space~\cite{Anastasiou:2014qba,Anastasiou:2018rdx,Luna:2020adi,Borsten:2019prq,Borsten:2020xbt,Luna:2020adi,Luna:2022dxo,Liang:2023zxo}. Given this, we note that the locality of the position-space double copy arises in this case because our $\phi$ is an eigenvector of the explicitly \emph{non-local} operator~${\hat V}_\mu$.
Enticingly, the same operator also appears in a classical double copy correspondence between the  \textit{off-shell} self-dual dyon and self-dual Taub-NUT~\cite{Kim:2024dxo}.
It would be interesting to investigate the extent to which this applies in other examples, as well as broader connections to non-locality in double copy~\cite{Emond:2025nxa}. 
%%%%%%%%%%%%%%%%%%%%%%%%%
\subsection{Double Kerr-Schild}
%%%%%%%%%%%%%%%%%%%%%%%%%
Finally, we show that the procedure of Sec.~\ref{sec:kerr-schild} has a natural extension beyond self-dual fields which connects to another familiar classical double copy structure.
Consider now a general gauge field written in a helicity basis as
\begin{equation}
    A_\mu(x) = \sfT \int_k  \bar\varepsilon^{s}_\mu {\bar \alpha}^{\prime}_s(k) \, e^{i k \cdot x} + \varepsilon^{s}_\mu  \alpha_s(k) \, e^{- i k \cdot x}\;.
\end{equation}
We can extract the positive helicity vectors using $\hat{V}_\mu$ as in (\ref{KSoperator}), and the negative  helicity vectors using its complex conjugate $\hat V_\mu^\dagger$, writing
\begin{equation}\label{DKS-gauge}
    A_\mu = \sfT \hat V_\mu \varphi_{+} + \sfT \hat{V}_\mu^\dagger \varphi_{-}  \;,
\end{equation}
compare~(\ref{single-copy}).  The scalar fields $\varphi_s$ have the same form as in~(\ref{KS-scalar}), giving us a linearised solution to the Yang-Mills equations (which may be real or complex depending on the choice of coherent state profiles). With our approach the two terms in (\ref{DKS-gauge}) double copy \emph{separately}, following precisely the procedure in Sec.~\ref{sec:kerr-schild}. The resulting metric is
\begin{equation}
    g_{\mu\nu} = \eta_{\mu\nu} + \kappa \hat V_\mu \hat V_\nu\varphi_{+}  + \kappa {\hat V}^\dagger_\mu {\hat V}_\nu^\dagger \varphi_{-},
\end{equation}
which is in double Kerr-Schild form, first discussed in a classical double copy context in~\cite{Luna:2015paa}. The construction used therein is based on performing the `standard' Kerr-Schild double copy \emph{independently on each term}; we now see that this choice arises naturally from applying double copy to amplitudes of coherent states. The double Kerr-Schild metric does not in general linearise Einstein's equations, and it is not as straightforward to identify vacuum solutions. Nevertheless, it is what our symmetric double copy scheme outputs for an arbitrary coherent gauge background based on helicity states.

%%%%%%%%%%%%%%%%%%%%%%%%%%%%%%%%%
\section{Conclusions}\label{sec:concs}
%%%%%%%%%%%%%%%%%%%%%%%%%%%%%%%%%
%
We have introduced a notion of double copy in which scattering amplitudes of coherent states in gauge theory double copy to amplitudes of coherent states in (axio-dilaton) gravity. Through the known equivalence between scattering of coherent states and scattering on background fields, we have shown that gauge theory amplitudes in any background admitting a coherent state description double copy to gravitational amplitudes on a background `fat metric' (meaning in curved spacetime, with a background axion and dilaton). Furthermore, the gauge and gravity backgrounds are related through a classical double copy relation inherited from double copy of amplitudes in vacuum. 
Though originating from the same replacement rules, we have seen that there is a neat separation between the classical double copy of backgrounds and that of the (spin-stripped) amplitudes. As such we may utilise the double copy to relax the computational burden of calculating the gravity amplitudes~(\ref{grav-amp-coherent-bgft}) in \textit{any} axio-dilaton background which admits a coherent state description, whether or not it has its roots in a classical double copy.

Due to the form of the coherent states used, the class of backgrounds we consider is only guaranteed to solve the linearised Maxwell and Einstein vacuum equations, and we have seen how existing results for plane waves fit into this scheme. We have also explored in detail the case of self-dual backgrounds, in which case our map relates \emph{exact} vacuum solutions in both gauge theory and gravity.

We used non-abelian coherent states which are constructed as products of coherent states for each colour -- this amounts to treating the asymptotic gluons in those states as free, just as is assumed when calculating ordinary amplitudes of number states.  Our approach was guided by the principle of using as little as possible beyond double copy relations for amplitudes in vacuum. The only condition we impose is that the chosen double copy scheme relates gauge and gravity amplitudes through~(\ref{n-point-DC}).

Turning to future work,  it would be interesting to see if our approach could be extended to \emph{off-shell} coherent states~\cite{Cristofoli:2020hnk,Adamo:2022ooq}, as this would allow us to consider sourced backgrounds, e.g.~shockwaves, Coulomb and Schwarzchild~\cite{Bahjat-Abbas:2020cyb,Easson:2020esh,Easson:2021asd,Armstrong-Williams:2024bog}.
Studying examples with horizons~\cite{Chawla:2023bsu,Ilderton:2023ifn,He:2023iew} would be particularly interesting.
Examining further examples of fields related by our  classical double copy could shed light on how gauge transformations and diffeomorphisms of classical backgrounds are related through double copy~\cite{Anastasiou:2014qba,Anastasiou:2018rdx}, and connections to other versions of double copy on e.g.~self-dual backgrounds~\cite{Adamo:2024hme}. Another possible direction would be to explore \emph{non}-perturbative extensions to our double copy map~\cite{Armstrong-Williams:2022apo}, asking if we can utilise the connections to background field theory to find an effective map which ``resums" perturbative double copy replacements. It seems from our examples and progress in the literature that a natural starting point for such an investigation would be the self-dual sector~\cite{Monteiro:2011pc,Boels:2013bi,Campiglia:2021srh,Armstrong-Williams:2022apo,Klisch:2025pbu,Monteiro:2022nqt,Correa:2024mub, Kim:2024mpy}.

\medskip
\textit{We thank Tim Adamo, Asaad Elkhidir and Donal O'Connell for useful discussions. The authors are supported by the STFC consolidated grant ``Particle Theory at the Higgs Centre" ST/X000494/1 (AI) and an STFC studentship (WL).}
\bibliography{refs}
\bibliographystyle{JHEP}

\end{document}